\documentclass[amsmath,amssymb,preprint]{revtex4}

\usepackage{graphicx}

\begin{document}

\title{Intermittent turbulence and turbulent structures in a linear magnetized plasma}
\author{T.A. Carter} \email{tcarter@physics.ucla.edu}
\affiliation{Department of Physics and Astronomy and Center for
  Multiscale Plasma Dynamics, University of California, Los Angeles,
  CA 90095-1547}

\begin{abstract} 

  Strongly intermittent turbulence is observed in the shadow of a
  limiter in the Large Plasma Device (LAPD) at UCLA [W. Gekelman,
    H. Pfister, Z. Lucky, J. Bamber, D. Leneman, and J. Maggs,
    Rev. Sci. Inst. {\bfseries 62}, 2875 (1991)].  The amplitude
  probability distribution function (PDF) of the turbulence is
  strongly skewed, with density depletion events (or ``holes'')
  dominant in the high density region and density enhancement events
  (or ``blobs'') dominant in the low density region.  Two-dimensional
  cross-conditional averaging shows that the blobs are detached,
  outward-propagating filamentary structures with a clear dipolar
  potential while the holes appear to be part of a more extended
  turbulent structure.  A statistical study of the blobs reveals a
  typical size of ten times the ion sound gyroradius and a typical
  velocity of one tenth the sound speed.

\end{abstract}


\maketitle

Turbulence which is intermittent, meaning patchy in space or bursty in
time, is often observed in both neutral fluids~\cite{frisch} and
plasmas~\cite{antar03}.  A signature of intermittency in turbulent
measurements is a non-Gaussian amplitude probability distribution
function (PDF), {\itshape e.g.} with tails caused by a greater
frequency of large amplitude events.  Intermittent turbulence is
ubiquitous in the edge of magnetic confinement laboratory plasmas
including tokamaks~\cite{boedo01}, stellarators~\cite{sanchez03}, and
linear devices~\cite{antar01}.  The intermittency in these
environments is generally attributed to the creation and propagation
of filamentary (magnetic field aligned)
structures~\cite{zweben85,krash01}.  Large amplitude events in
Langmuir probe~\cite{rudakov02}, beam emission
spectroscopy~\cite{boedo03}, and gas puff imaging
measurements~\cite{zweben02} are therefore due to the passage of high
density structures through the low density edge region. The
outward propagation of these structures results in significant
particle transport in the edge of magnetic confinement devices like
tokamaks~\cite{dippolito02}.  In addition, recently a correlation has
been found between increased intermittency in the scrape-off-layer
(SOL) and the density limit in tokamaks, leading some to suggest that
catastrophic transport enhancements associated with intermittent
turbulence may be responsible for this disruptive
limit~\cite{greenwald02}.

Several mechanisms have been proposed for the generation of these
structures including solitary drift wave
vortices~\cite{hassam79,horton89}, propagating avalanche-like events
in the plasma core~\cite{diamond95,newman96}, zonal-flow driven
generation~\cite{smolyakov00} and interchange driven
production~\cite{garcia04}.  Once the structures are produced and
ejected into the low density region, their continued cross-field
propagation has been attributed to $E\times B$ velocity due to
polarization by drift-charging~\cite{chen67,krash01}.  In tokamaks,
the drift-charging is attributed to interchange forces associated with
magnetic field gradients and curvature~\cite{krash01}.  In linear
devices, rotation~\cite{antar03} is usually invoked as an interchange
force.

In this Letter, a detailed study of turbulent structures associated
with intermittent turbulence in a linear magnetized plasma with a
steep density gradient is presented. In this study, a clear
observation of both density enhancements (or ``blobs'') and density
depletions (or ``holes'') is made.  The production of blob and hole
structures is consistent with an interchange-like process.  However,
the experiments occur in a plasma with straight, uniform magnetic
field and without significant rotation due to the straight, vertical
edge created by the limiter. The typical interchange forces are
therefore not available, yet the resulting intermittency is very
similar to that observed in tokamaks.  It has been proposed that a
frictional force due to neutral flows may serve as an interchange
force in cases where rotation or curvature are not
present~\cite{krash03}, and this theory may explain the observations. 
Two-dimensional measurements of cross-field structure show that blobs
are isolated structures with a dipolar potential while the holes are
instead part of a more extended structure.  A statistical study of the
properties of the blobs has been performed, showing that the size of
the structures scales like the ion sound gyroradius ($\delta \sim
10\rho_s$) and that the velocity of the objects (outward, into the
limiter shadow) scales with the sound speed ($v \sim C_s/10$).

\begin{figure}[!htbp]
\centerline{
\includegraphics[width=3.6truein]{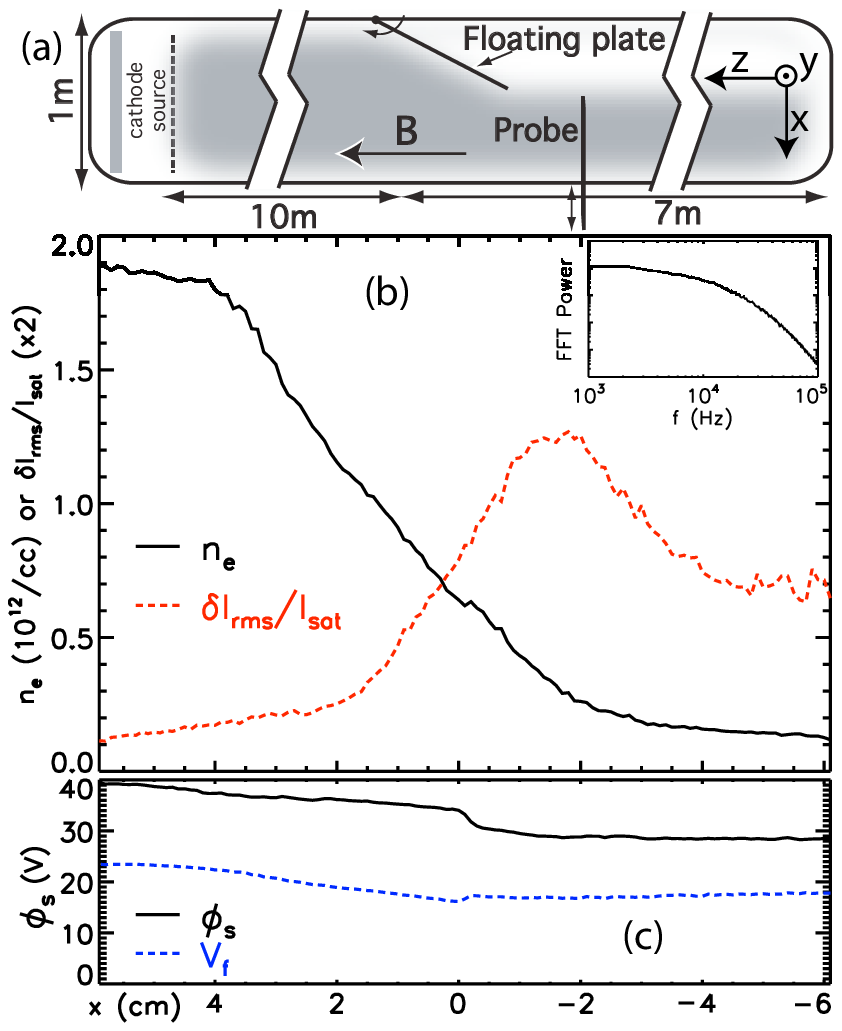}}
\caption{(Color online) (a) A schematic of the LAPD, showing the experimental
  geometry. (b) Spatial profile of density and fluctuations behind the
  limiter ($x=0$ at the limiter edge), along with an average
  fluctuation power spectrum (log-log). (c) Spatial profile of space
  potential and floating potential, as measured using a triple
  probe.}\label{figure1}
\end{figure}

The experiments were performed in the upgraded Large Plasma Device
(LAPD), which is part of the Basic Plasma Science Facility
(BaPSF)~\cite{lapd} at UCLA. LAPD is an 18m long, 1m diameter
cylindrical vacuum chamber, surrounded by 90 magnetic field
coils. Pulsed plasmas ($\sim 10$ms in duration) are created at a
repetition rate of 1Hz using a Barium Oxide coated cathode
source. Typical plasma parameters are $n_e \sim 1 \times 10^{12}$
cm$^{-3}$, $T_e \sim 5$eV, $T_i \sim 1$eV, and $B < 2$kG.  The
working gases used in these experiments were Helium and Neon. The
length of the LAPD plasma column can be changed by closing a hinged,
electrically-floating aluminum plate located 10m from the cathode.
In the experiments reported here, this plate is partially closed and
used to limit the plasma column as shown in
Figure~\ref{figure1}(a). With the plate partially closed, there is no
parallel (to the magnetic field) source of plasma for the region
behind the plate and any observed plasma density must come from
cross-field transport.  Measurements were performed using
radially-insertable Langmuir probes [as shown in
Fig.~\ref{figure1}(a)].  Triple Langmuir probes with 2mm long,
0.76mm diameter cylindrical tungsten tips spaced 3mm apart were used
to measure density, electron temperature and floating potential and to
perform two-dimensional cross-conditional averaging measurements.  A
linear array of six single Langmuir probes (1.5mm diameter
flush-mounted tantalum tips spaced 5mm apart, aligned with their surface normals along
the magnetic field) was also used to study the propagation of structures.

For these experiments, the floating-plate limiter was closed so that
nearly half of the plasma column is blocked, and a half-cylindrical
plasma is observed downstream from the limiter.  Steep gradients in
plasma density are observed behind the limiter, along with very large
amplitude ($\delta n/n \sim 1$) fluctuations. Figure~\ref{figure1}(b)
shows a measurement of the radial profile of plasma density ($n_{\rm
  e}$) and normalized root-mean-square (RMS) ion saturation current
($I_{\rm sat}$) fluctuation amplitude ($\delta I_{\rm rms}/I_{\rm
  sat}$) 0.5m downstream from the limiter (He discharge, 1.5kG field)
along with an average fast Fourier transform (FFT) power spectrum of
the fluctuations.  The observed density gradient is very steep with a
scale length of order a few centimeters (the ion sound radius, $\rho_s
\sim 0.3$cm and the ion gyroradius, $\rho_i \sim 0.15$cm).  In the
limiter shadow, the density profile is flat, similar to the profiles
observed in the scrape-off-layer of tokamaks~\cite{boedo03}. The
density behind the limiter is a significant fraction of the core
density, indicative of substantial cross-field particle transport.
The fast Fourier transform (FFT) power spectrum of the fluctuations is
broadband and free from coherent modes.  Figure~\ref{figure1}(c) shows
the profile of space potential (derived from floating potential, also
shown) in the plasma.  While there is a strong feature in the
potential profile near the location of the limiter, indicating a
sheared $E\times B$ flow, there is little or no flow in the region
behind the limiter.  It is important to note that the limiter has a
straight vertical edge. Thus, while there are vertical flows localized
to the limiter edge region, there is no rotation and associated
centrifugal force.

\begin{figure}[!htbp]
\centerline{
\includegraphics[width=3.6truein]{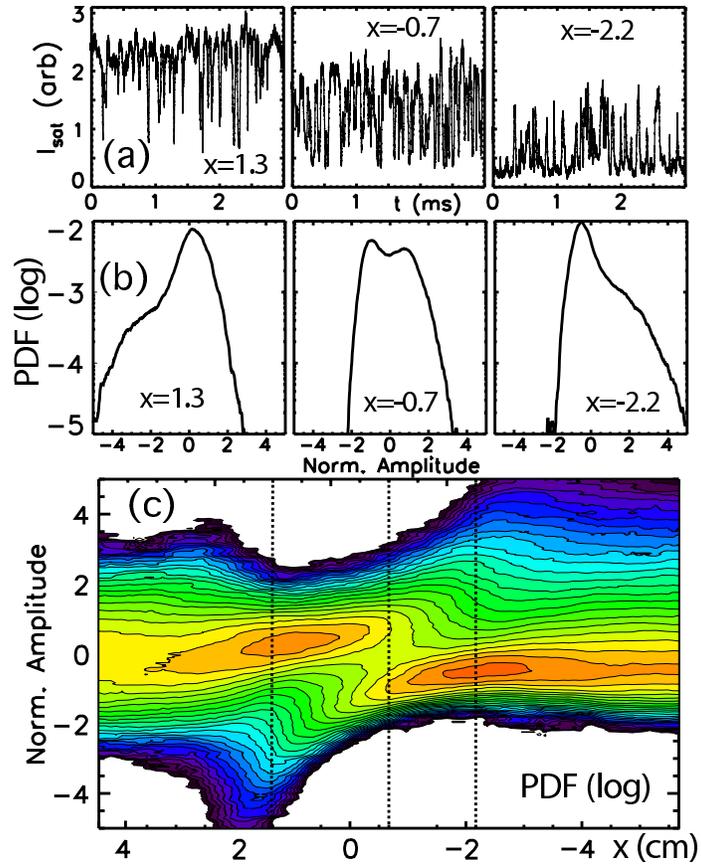}}
\caption{(Color online) (a) Example raw $I_{\rm sat}$ signals at
  three spatial locations.  (b) Amplitude PDF at
  the same three locations. (c) Contour plot of the amplitude PDF
  versus position. }\label{figure2}
\end{figure}

Figure~\ref{figure2}(a) shows example $I_{\rm sat}$ signals
measured at three spatial locations ($x=1.3, -0.7, -2.2$cm, where
$x=0$cm is the edge of the limiter).  On the core side of the
gradient region ($x=1.3$cm), the signal is dominated by
downward-going events while on the low-density side of the gradient
region ($x=-2.2$cm) the signal is dominated by upward-going events.
The upward-going events in the low density region are density
enhancements or ``blobs'', similar to those observed in the edge of
many magnetic confinement devices~\cite{antar03}.  The downward-going
events are identified as density depletions or ``holes,'' which are
not typically observed in other devices, but have been reported to be
seen in measurements on the DIII-D tokamak~\cite{boedo03}.  Generation
of holes has been observed in simulations~\cite{russell04} and inward
propagation of holes has been suggested as a mechanism for impurity
transport in tokamaks~\cite{dippolito04}.  Although the raw signals
clearly show the intermittency in the fluctuations, a more
quantitative measurement of the intermittency can be derived from the
fluctuation amplitude PDF.  The amplitude PDF has been computed as a
function of the amplitude normalized to the root-mean-square (RMS)
fluctuation level, and is shown for $x=1.3, -0.7, $and $-2.2$~cm in
figure~\ref{figure2}(b).  The existence of the holes and blobs
produces tails in the PDF at negative and positive normalized
amplitude, respectively.  The PDF at the location of peak fluctuation
amplitude ($x=-0.7$cm) is more symmetric, although is still distinctly
non-Gaussian. The amplitude PDF as a function of position is shown as
a contour plot in figure~\ref{figure2}(c).  From this figure it is
evident that the holes are observed only in a narrow region of space
on the high density side of the gradient (where the negative tail is
present, $2.8 \gtrsim x \gtrsim 0$).  However the blobs are observed
everywhere to the right of the density gradient region ($x \lesssim
-0.5$), as shown by the presence of a tail in the PDF at positive
amplitude.  For $x \lesssim -3$cm, the PDF is blob-dominated and
nearly independent of position, suggesting that the blob structures
are long lived as they propagate into the shadow of the limiter
(intermittent signals have been observed up to 20cm beyond the limiter
edge).  It should be noted that the PDF displayed in
figure~\ref{figure2}(c) exhibits trends which are very similar to
those observed using BES in the DIII-D tokamak (see figure 3 of
reference 9).  The nature of the intermittency and turbulent
structures in these experiments is therefore quite similar to that
observed in toroidal confinement devices, even in the absence of
magnetic curvature or rotation.

\begin{figure}[!htpb]
\centerline{
\includegraphics[width=3.6truein]{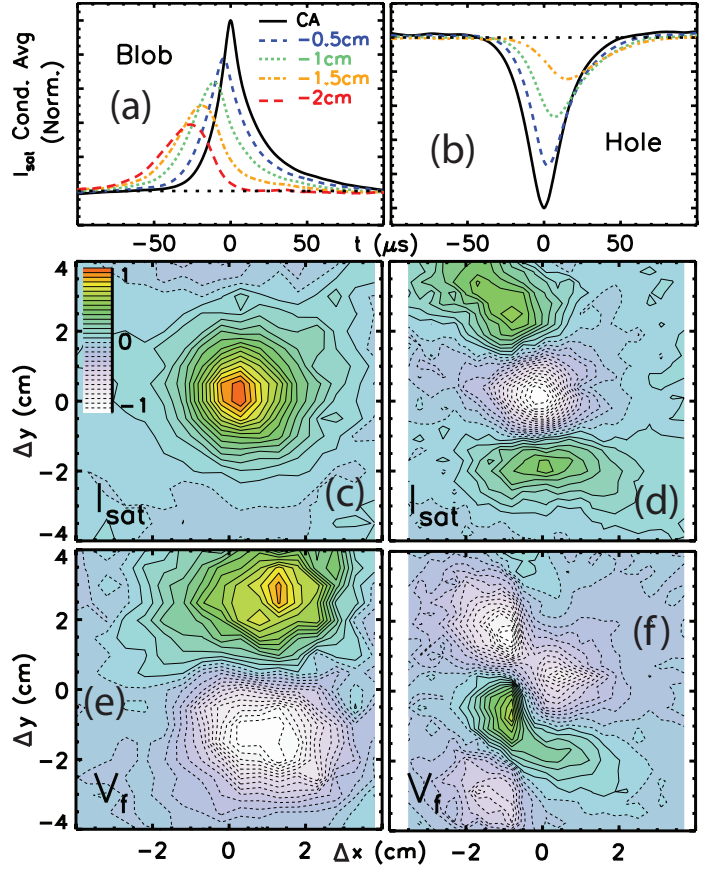}}
\caption{(Color online) Cross-conditional average of $I_{\rm sat}$ on the linear
  probe array for (a) blob and (b) hole events, showing apparent
  propagation of blobs out of the plasma and holes back into the
  plasma.  Two dimensional cross-conditional averages of blob ((c)
  $I_{\rm sat}$ and (e) $V_{\rm f}$) and hole ((d) $I_{\rm sat}$ and
  (e) $V_{\rm f}$).  All 2D conditional averages are normalized to the
  maximum of the absolute value of the average, and the color bar in
  (c) applies to all images.}\label{figure3}
\end{figure}

Figure~\ref{figure3}(a) shows the conditional average~\cite{filipas95}
of many ($N=16059$) blob events using $I_{\rm sat}$ signals
from one tip on the linear Langmuir array (solid black line, marked
``CA''). The amplitude threshold for triggering event selection in
this case was set to twice the root-mean-square fluctuation amplitude.
The conditional average blob event is asymmetric in time, with a fast
rise and slow decay.  This time asymmetry is consistent with
observations in the edge of many other magnetic confinement
devices~\cite{antar03} and with simulations~\cite{dippolito04}.
Along with the conditional average, figure~\ref{figure3}(a) shows the
cross-conditional average blob event as measured by other tips in the
Langmuir array (separations shown are in the $-\hat{x}$ direction).
The right-most tip (furthest away from the gradient region in the blob
case) is used for event triggering.  The average blob appears to
travel across the probe array out into the low density region with a
speed of $\sim 942$m/s, which is $\sim C_{\rm s}/10$.
Figure~\ref{figure3}(b) shows the same conditional and
cross-conditional averaging for hole events.  The average hole decays
quickly in space, but appears to propagate back into the core plasma.
The interpretation of the Langmuir array conditional averages as being
caused by propagation of structures in and out of the core plasma can
not be conclusive without knowledge of the two-dimensional structure
of the objects.  For example, the observations could also be explained
by, {\itshape e.g.}, the vertical ($\hat{y}$) propagation of
structures which are tilted in the $xy$ plane.  

In order to conclusively determine the structure and direction of
propagation of the holes and blobs, two-dimensional cross-conditional
averaging was performed using two triple Langmuir probes separated
along the magnetic field by 60~cm.  The first probe, which was the
reference or trigger probe, was left fixed in space while the second
probe was moved to 441 positions in a 10cm by 10cm cross-field ($xy$)
plane centered on the position of the reference probe.
Figure~\ref{figure3}(c) and (e) show the 2D cross-conditional average
of $I_{\rm sat}$ and $V_{\rm f}$, respectively, for blob events (the
fixed probe was located at $x=-3$cm).  The blob is clearly an
isolated, detached structure, with similar extent in the two
cross-field directions.  The $V_{\rm f}$ measurement clearly shows a
dipole structure of the potential associated with the blob (the0 peak
potential value is $\sim 1.3$V while the minimum value is $\sim
-1.5V$, here $T_e \sim 5$eV).  The potential structure is consistent
with $E\times B$ propagation almost entirely in the $-\hat{x}$
direction with an average speed of $\sim 985$m/s.  This value is
comparable to the speed measured by the linear probe array.  Because
the blob velocity is dominantly in the $-\hat{x}$ direction, the decay of
the cross-conditional average in figure~\ref{figure3}(a) can be
attributed to a finite spread in the blob
velocity. Figure~\ref{figure3}(d) and (f) show 2D cross-conditional
averages for hole events (fixed probe at $x = 1.5$cm).  The hole does
not appear to be an isolated structure, but instead is associated with
turbulent structure that is extended in the vertical ($\hat{y}$)
direction.  The potential structure does show a tendency for
$E\times B$ propagation back into the core plasma, consistent with the
linear array measurements.

\begin{figure}[!htpb]
\centerline{
\includegraphics[width=3.6truein]{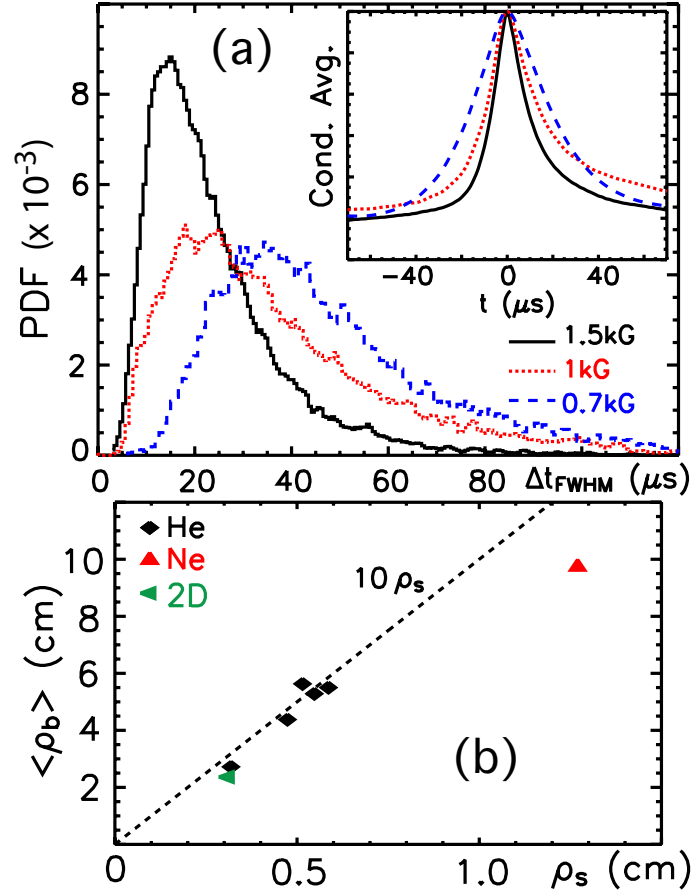}}
\caption{(Color online) (a) PDF of blob event time width, for three values of the
  magnetic field.  Inset are conditional average blob events for the
  three magnetic field values. (b) Mean blob size for several
  conditions in He and Ne, compared to the ion sound gyroradius.}\label{figure4}
\end{figure}

Figure~\ref{figure4}(a) shows the measured dependence of the time
width of blob events versus magnetic field.  The inset figure shows
the conditional average $I_{\rm sat}$ signal for a blob event for
three field values.  As the field is decreased, the time width of the
event increases.  The main figure shows the PDF of the time width
[full width at half max (FWHM)] of blob events for the same three
magnetic field values.  To calculate the PDF of the blob time width,
events are selected from the $I_{\rm sat}$ signals and the
time width ($\Delta t_{\rm FWHM}$) of each individual event is
measured.  As the magnetic field is decreased, the peak of the PDF
shifts to larger time width and also the width of the PDF increases.
The width in time of the blob event can be translated to a blob size
using the blob velocity as measured by the linear probe array.
Figure~\ref{figure4}(b) shows the average blob size ($\left<\rho_{\rm
  b}\right>$) compared to the ion sound gyroradius ($\rho_{\rm s}$)
for blob events measured in several different magnetic field values in
Helium and one condition using Neon as the working gas. In addition,
the average size obtained from the 1.5kG 2D cross-conditional average
[figure~\ref{figure3}(c)] is shown. The average blob size scales with
the ion sound gyroradius, and is approximately ten times this scale.
The observation of a gyroradius scaling is consistent with recent
theory~\cite{dippolito04a} which attributes the scaling to blob
stability.  It should also be noted that the average blob size is also
comparable to the density gradient scale length.

In summary, both blobs and holes are observed to be associated with
strongly intermittent turbulence in the shadow of a limiter in LAPD.
The blobs are nearly cylindrical filamentary structures which have a
dipolar potential, resulting in $E\times B$ propagation into the low
density limiter shadow.  A statistical study of the blobs has been
performed, showing that the average speed of the structures is $\sim
C_{\rm s}/10$ and that the average blob size is $\sim 10 \rho_{\rm
  s}$.  The holes do not appear to be detached, propagating objects,
but are instead part of a more extended turbulent structure, a finding
which might also apply to observations of holes in tokamaks. These
observations are made in an experiment free from traditional
interchange forces, yet are very similar to measurements in the edge
of toroidal confinement devices.  The formation of these turbulent
structures and their polarization may be linked to the instabilities
associated with sheared flow in the gradient region.  The dipolar
potential of the blobs may also be explained by drift-charging arising
from a newly-proposed force due to neutral flows.  An initial
calculation using this theory and LAPD parameters results in a blob
velocity which agrees with the experimental
observation~\cite{krash03}.  The role of sheared flows and neutrals
will be explored in more detail in future research.

Discussions with G. Antar, S. Krasheninnikov, R. Moyer, A. Pigarov and
D. Rudakov are gratefully acknowledged.  The experiments reported in
the paper were performed using the UCLA Basic Plasma Science Facility,
which is funded by NSF and DOE.  This work was supported by
a DOE Fusion Energy Sciences postdoctoral fellowship and by
DOE grant DE-FG02-02ER54688.

\newpage

\end{document}